\documentclass[prl,twocolumn,showpacs]{revtex4-1}

\usepackage{amsmath,amssymb}
\usepackage{graphicx}
\usepackage{color}
\usepackage{dcolumn}

\newcommand{\be}{\begin{equation}}
\newcommand{\ee}{\end{equation}}
\newcommand{\bea}{\begin{eqnarray}}
\newcommand{\eea}{\end{eqnarray}}
\newcommand{\nn}{\nonumber}
\newcommand{\eins}{\leavevmode\hbox{\small1\kern-3.8pt\normalsize1}}
\newcommand{\e}{\mbox{e}}

\newcommand{\ga}{\gamma}

\newcommand{\Tr}{\mbox{Tr}}
\newcommand{\Pf}{\mbox{Pf}}

\begin{document}

\title{
Completing the picture for the smallest eigenvalue of real Wishart matrices
}

\author{G. Akemann$^1$, T. Guhr$^2$, M. Kieburg$^1$, R. Wegner$^1$, and T. Wirtz$^2$} 

\affiliation{$^1$ Fakult\"at f\"ur Physik, Universit\"at Bielefeld, 
D-33501 Bielefeld, Germany\\
$^2$Fakult\"at f\"ur Physik, Universit\"at Duisburg-Essen, D-47048 Duisburg, Germany
}

\date{\today}

\begin{abstract}
  Rectangular real $N \times (N + \nu)$ matrices $W$ with a Gaussian
  distribution appear very frequently in data analysis, condensed
  matter physics and quantum field theory. A central question concerns
  the correlations encoded in the spectral statistics of $WW^T$. The
  extreme eigenvalues of $W W^T$ are of particular interest. We
  explicitly compute the distribution and the gap probability of
  the smallest non-zero eigenvalue in this ensemble, both for
  arbitrary fixed $N$ and $\nu$, and in the universal large $N$ limit
  with $\nu$ fixed. We uncover an integrable Pfaffian structure valid
  for all even values of $\nu\geq 0$. This extends previous results
  for odd $\nu$ at infinite $N$ and recursive results for finite $N$
  and for all $\nu$. Our mathematical results include the computation
  of expectation values of half integer powers of characteristic
  polynomials.

\end{abstract}



\maketitle

\textbf{Introduction.} To study generic statistical features of
spectra, various kinds of random matrices are used. Following Wigner
and Dyson~\cite{Dyson1962}, 
Hamiltonians of
dynamical systems are modelled by real--symmetric, Hermitian or self-dual 
matrices in quantum chaos, many-body and
mesoscopic physics. Due to universality, cf. ~\cite{Guhr1998189,handbook} 
and references therein, Gaussian probability densities
suffice, leading to the Gaussian Orthogonal, Unitary and Symplectic
Ensemble (GOE, GUE, GSE)~\cite{Mehta}. This concept was
extended
to Dirac spectra~\cite{ShuryakVerbaarschot}  
by
imposing chiral symmetry as an additional constraint, resulting in the
chiral ensembles chGOE, chGUE, chGSE~\cite{Verbaarschot1994}.  Wishart~\cite{Wishart} put
forward random matrices to model spectra of correlation matrices in a quite different context.  There are 
many
applications in time series analysis
\cite{Muirhead,Anderson,Chatfield}
(including chaotic dynamics \cite{VP}), 
in a wide range of fields in
physics~\cite{Guhr1998189,handbook}, biology \cite{Seba}, wireless communication \cite{TV} and
finance \cite{Plerou}.  In the most relevant case, $N \times (N +
\nu)$ real matrices $W$ model 
time series such that $WW^T$ is the
random correlation matrix. If it fluctuates around a given
average correlation matrix $C$, the distribution reads 
\be 
{\mathbb
  P}_{N,\nu}(W|C) \sim \exp\left[-
  \Tr WW^T C^{-1}/2\right] \ .
\label{measure}
\ee For $C=\eins_N$, this happens to coincide with the chGOE, where
$W$ and $W^T$  model the non--zero blocks of the Dirac
operator. Closing the circle, one can also extend Wishart's model by using
non-Gaussian weights. Here and in the sequel, we focus on Eq.~\eqref{measure} with
$C=\eins_N$. Since $WW^T$ has positive eigenvalues, the spectrum is bounded from
below. Naturally, the distribution of the smallest (non--zero)
eigenvalue is of particular importance.

Much interest in the chGOE was sparked by the observation \cite{Jac}
that in the limit $N\to\infty$ its spectral correlators describe the
Dirac spectrum in quantum field theories with real Fermions and broken
chiral symmetry, see \cite{Verbaarschot:2000} for a review.  Based on
earlier works for finite $N$ \cite{Mahoux,NW}, the spectral density
\cite{Jac} and all higher density correlation functions \cite{NF95}
were computed in terms of a Pfaffian determinant of a matrix kernel
for all $\nu$. These quantities were shown later to be universal
\cite{Klein} for non-Gaussian potentials, and most recently for fixed
trace ensembles in the context of quantum entanglement, see \cite{AV}
and references therein.  Further applications of the chGOE can be
found in the recent review \cite{carlo} on Majorana Fermions and
topological superconductors.

In an influential paper \cite{Edelman88} the condition number of a
Wishart random matrix $WW^T$ was investigated, that is the root of 
the ratio of the largest over the smallest non-zero eigenvalue of $WW^T$. 
This quantity is important for a generic matrix as it quantifies the 
difficulty of computing its inverse. In
\cite{Edelman} the distribution of the smallest eigenvalue was
calculated recursively in $N$ for arbitrary rectangular chGOE
matrices. Closed expressions were given for quadratic matrices $\nu=0$
\cite{Edelman88} (cf. \cite{Peter93}) and for $\nu=1,2,3$
\cite{Edelman}. Later Pfaffian expressions were found in \cite{NF98}
for arbitrary odd $\nu$ valid for fixed and asymptotically large
$N$. A more general consideration, including correlations with $C\neq \eins_N$, 
of the smallest eigenvalue for $\nu$ odd
was given in \cite{WG}.  The limiting distributions of the $k$-th
smallest eigenvalue were computed in \cite{DN01}, again
for $\nu$ odd.  These quantities are an efficient tool to test
algorithms with exact chiral symmetry in lattice gauge theories
\cite{Edwards}, distinguishing clearly between different topologies
labelled by $\nu$.  In \cite{BLP} the distributions for higher even
$\nu>0$ were obtained from numerical chGOE simulations.  Most recently
efficient numerical algorithms have been applied, see e.g. \cite{SMN},
in order to compute smallest eigenvalue distributions for arbitrary
$\nu$ using known analytic Fredholm determinant expressions
\cite{Peter06}.

It is our goal to complete the picture for the smallest chGOE
eigenvalue distribution and its integral by finding explicit Pfaffian
expressions for finite and infinite $N$ valid for all even $\nu$.
Together with previous results this completes the integrability of
this classical ensemble. A presentation with further results and more
mathematical details will be given elsewhere \cite{AKWW}.


\textbf{Smallest eigenvalue and gap probability.}  
First we define the quantities of interest and state the problem.
In the analytic calculations below we set $C=\eins_N$ in Eq.~(\ref{measure}),
and later we compare our universal large $N$ results to 
numerical simulations with $C\neq\eins_N$.
Because we are only interested in correlations of the positive eigenvalues of $WW^T=OXO^T$ contained in $X=\mbox{diag}(x_1,\ldots,x_N)$, we drop all normalisation constants depending on the orthogonal matrix $O$.
Integrating the distribution (\ref{measure}) over all independent matrix elements with respect to flat Lebesgue measure we obtain the partition function expressed in terms of the eigenvalues as 
\be
{\cal Z}_{N,\gamma}=\prod_{i=1}^N\int_0^\infty dx_iw_\ga(x_i)\ |\Delta_N(X)|\ ,
\label{Z}
\ee
up to a known constant. Here we introduce the weight function $w_\ga(x)$ and 
Vandermonde determinant $\Delta_N(X)$ stemming from the Jacobian of the diagonalisation,
\bea
w_\ga(x)&\equiv& x^\gamma\exp[-x/2]\ , \ \gamma\equiv{(\nu-1)/2}\ ,
\label{weight}\\
\Delta_N(X)&\equiv&\prod_{1\leq i<j\leq N}(x_j-x_i)
=\det_{1\leq i,j\leq N}\left[x_i^{j-1}\right].
\label{Vandermonde}
\eea
We note that $\gamma$ alternates between integer and half-integer values.
The expectation value of an observable $f$ only depending on $X$ is defined as 
\be
\langle f(X)\, \rangle_{N,\ga}\equiv 
\frac{\prod_{i=1}^N\int_0^\infty dx_iw_\ga(x_i)\ f(X)|\Delta_N(X)|}{{\cal Z}_{N,\gamma}}.
\label{vev}
\ee
Thus the gap probability that no eigenvalue occupies the interval $[0,t]$ is given by
\bea
{ E}_{N,\ga}(t)&\equiv& 
\frac{1}{{\cal Z}_{N,\gamma}}
\prod_{i=1}^N\int_t^\infty dx_iw_\ga(x_i)\ |\Delta_N(X)|
\nn\\
&=&\e^{-Nt/2}\frac{{\cal Z}_{N,0}}{{\cal Z}_{N,\gamma}}
\left\langle {\det}^\ga[X+t\eins_N]\right\rangle_{N,0}\ .
\label{charP}
\eea
It is expressed as an expectation value of 
a
characteristic polynomial 
to the 
power $\ga$ with respect to the weight function (\ref{weight}) without the pre-exponential factor, $w_0(x)$.
This crucial identity 
follows from the translation invariance of the Vandermonde determinant
(\ref{Vandermonde}).
 
The normalised distribution of the smallest non-zero eigenvalue, ${ P}_{N,\ga}(t)$, 
is obtained by differentiating Eq.~(\ref{charP}) 
\bea
{ P}_{N,\ga}(t)&\equiv&-\frac{\partial{ E}_{N,\ga}(t)}{\partial t} 
\label{Pmin}\\ 
&=&t^\ga \e^{-{Nt}/{2}}\frac{N{\cal Z}_{N-1,1}}{{\cal Z}_{N,\gamma}}
\left\langle {\det}^\ga[X+t\eins_{N-1}]\right\rangle_{N-1,1},\nn
\eea 
where the second line follows along the same steps as in Eq.~(\ref{charP}). This relation is well known \cite{NF98,DN01}, with the difficulty to compute the average (also called massive partition function) for $\ga$ half-integer, which is our main task.

To compute Eqs.~(\ref{charP}) and (\ref{Pmin}) we need to know the normalising partition functions, which are given for arbitrary real $\nu>-1$ in terms of the Selberg integral, see also \cite{NN,AK}, and the expectation values. For integer $\ga=k$ corresponding to odd $\nu=2k+1$ closed 
expressions of (\ref{Pmin}) exist \cite{NF98}, 
given in terms of Laguerre polynomials skew-orthogonal 
with respect to the weight (\ref{weight}). Therefore we concentrate on the case $\nu=2k$ even.

\textbf{Pfaffian structure and finite $N$ results.}
To show that the gap probability (\ref{charP}) has a Pfaffian structure when $\gamma$ is half-integer let us define the following parameter dependent weight function
\be
w(x;t)\equiv \exp[-\eta x/2]/\sqrt{x+t}\ .
\label{sqrtweight}
\ee
It absorbs the half-integer part in the expectation value (\ref{charP}) when $\nu=2k$ is even. We set $\eta=1$ unless otherwise stated.
The monic polynomials $R_k(x;t)=x^k+\ldots$ are defined to be skew-orthogonal with respect to
the following skew-symmetric scalar product
\be
\left<f,g\right>_t \equiv \int\limits^\infty_0\!\!dy\int\limits_0^y\!\!dx\, w(x;t)w(y;t)[f(x)g(y)-f(y)g(x)]
\label{scalarprod}
\ee
by satisfying for all $i,j=0,1,\ldots$~\cite{footnote1} 
the conditions
\bea
\left<R_{2j},R_{2i}\right>_{t} &=& 0 = \left<R_{2j+1},R_{2i+1}\right>_{t}\ \nn\\
\left<R_{2j+1},R_{2i}\right>_{t} &=& r_j(t)\ \delta_{ij}\ .
\label{skewdef}
\eea
Their normalisations $r_j(t)$ depend on $t$. 
The partition function ${\cal Z}_{N}(t)$ of this new weight (\ref{sqrtweight}) is defined by 
\be
{\cal Z}_{N}(t)\equiv\prod_{i=1}^N\int_0^\infty dx_iw(x_i;t)|\Delta_N(X)|
=
N!\prod_{i=0}^{\frac{N}{2}-1} r_i(t).
\label{Znew}
\ee
The last step holds for $N$ even \cite{Mehta}. 
Likewise we define expectation values 
$\langle f(X)\rangle_N^t$, 
following Eq.~(\ref{vev}). Thus for even $\nu=2k$, $k\in\mathbb N$, 
Eq.~(\ref{charP}) reduces to 
\be{ E}_{N,k-\frac12}(t)=
\e^{-{Nt}/{2} }
\frac{{\cal Z}_{N}(t)}{{\cal Z}_{N,k-\frac12}}
\left\langle {\det}^k[X+t\eins_N]\right\rangle_{N}^t,
\label{charPs}
\ee
given in terms of 
an
integer power of 
a
characteristic polynomial.
While the skew-orthogonal polynomials with respect to the weight (\ref{weight}) are know in terms of Laguerre polynomials \cite{NF98},
the difficulty here is to determine the $t$-dependent polynomials and normalisation constants for the non-standard weight (\ref{sqrtweight}). They can be computed following the observation \cite{Eynard}
\bea
R_{2j}(y,t)&=& 
\left\langle {\det}[X-y\eins_{2j}]\right\rangle_{2j}^t\ ,
\label{Reven}\\
R_{2j+1}(y,t)&=& 
\left\langle (y+c+\Tr X){\det}[X-y\eins_{2j}]\right\rangle_{2j}^t\nn\\
&=& (y+c) R_{2j}(y,t)- 2\left.\frac{\partial}{\partial\eta}R_{2j}(y,t)\right|_{\eta=1}\!\!\!\!\!.
\ \ \ \ \ 
\label{Rodd}
\eea
The odd polynomials are obtained by
differentiation of the weight (\ref{sqrtweight}), generating $\Tr X$ in the average. 
Note that the $R_{2j+1}(y,t)$
are not unique \cite{Eynard}, we set $c=0$ in the following.
The even polynomials (\ref{Reven}) can be calculated by mapping them back to a proper matrix integral over an auxiliary $2j\times(2j+1)$ matrix $\overline{W}$ (corresponding to $\gamma=0$),
\be
\label{R2jrep}
R_{2j}(y,t)
=C_{2j}(t)
\int d\overline{W}\frac{\det[\overline{W}\overline{W}^T-y\eins_{2j}]}{\det^{\frac12}[\overline{W}\overline{W}^T+t\eins_{2j}]}\e^{-\frac\eta2\Tr\overline{W}\overline{W}^T},
\ee
cf. \cite{WG}. 
The known normalisation constant $C_{2j}(t)$ follows from the fact that the polynomial is monic.
Without giving details Eq.~(\ref{R2jrep}) can be computed exactly, representing the determinants by Gaussian integrals over commuting and anti-commuting variables and by using standard bosonisation techniques
\cite{SW}.
We arrive at 
\be
\label{eq:DifferentiatedEvenPolynomials}
R^{a}_{2j}(y,t)=\frac{(2j)!\left(\text{U}_j(t){L}^{(a+1)}_{2j-a}(y)
-\text{U}^\prime_j(t)
{L}^{(a)}_{2j-a}(y)\right)
}{(2j-a)!\text{U}\left(\frac{2j+1}{2},\frac{3}{2},\frac{t}{2}\right)} 
\ee
for the $a$-th derivatives of the polynomials, $\frac{\partial^a}{\partial y^{a}}R_{j}(y,t)\equiv R^{a}_{j}(y,t)$, $a=0,1,\ldots$ needed later. Here $\text{U}_j(t)\equiv\text{U}\left(\frac{2j+1}{•2},\frac12,\frac{t}{2}\right)$ denotes the Tricomi confluent hypergeometric function, 
satisfying
$\text{U}^\prime(a,b,t)=-a\text{U}(a+1,b+1,t)$  \cite{NIST}.
The derivative in Eq.~(\ref{eq:DifferentiatedEvenPolynomials}) acts only on the generalised Laguerre polynomials used in {\it} {\it monic normalisation} ${L}^{(a)}_{j}(y)=y^j+\ldots$ They satisfy $\frac{\partial^a}{\partial y^{a}}
{L}^{(b)}_{n}(y)=\frac{n!}{(n-a)!}{L}^{(b+a)}_{n-a}(y)$, where we set 
${L}^{(b)}_{n}(y)\equiv0$ for $n<0$. 
For the odd polynomials we obtain
\bea
&&R^{a}_{2j+1}(y,t)=\left(4j^2+4j+y\right)R^{a}_{2j}(y,t)
+aR^{a-1}_{2j}(y,t)
\nn\\
&&+\frac{(2j)!/(2j-a)!}
{\text{U}\left(\frac{2N+1}{2},\frac{3}{2},\frac{t}{2}\right)}
\left\{t\text{U}^{\prime\prime}_j(t){L}^{(a)}_{2j-a}(y)
+2\text{U}^{\prime}_j(t)\right.\nn\\
&&
\times\left[ a{L}^{(a)}_{2j-a}(y)
+(2j-a)y{L}^{(a+1)}_{2j-a-1}(y)+\tfrac{t}{2}{L}^{(a+1)}_{2j-a}(y)\right]
\nn\\
&&\left.-2\text{U}_j(t)
\left[ a{L}^{(a+1)}_{2j-a}(y)+(2j-a)y{L}^{(a+2)}_{2j-a-1}(y)\right]
\right\},
\label{eq:DifferentiatedOddPolynomials}
\eea
and the normalisation constants in Eq.~(\ref{skewdef}) read
\begin{equation}
r_{j}(t) =2 (2j)!(2j+1)! \frac{\text{U}\left(\frac{2j+3}{2},\frac{3}{2},\frac{t}{2}\right)}{\text{U}\left(\frac{2j+1}{2},\frac{3}{2},\frac{t}{2}\right)}~.
\label{eq:valuescalarproduct}
\end{equation}
Following \cite{NF98} with their Laguerre weight $w_0(x)$ in Eq.~(\ref{weight}) replaced by our weight (\ref{sqrtweight}), we express the gap probability (\ref{charPs}) as a Pfaffian determinant with 
our kernel consisting of the skew-orthogonal polynomials (\ref{eq:DifferentiatedEvenPolynomials}) and (\ref{eq:DifferentiatedOddPolynomials}).
In a more general setting averages of characteristic polynomials such as Eq.~(\ref{charPs})
were considered in Refs. \cite{BS,KG} for arbitrary but unspecified weights. 
For finite even $N$ and $\nu = 2k$ with $k = 2m$ even we obtain
\begin{eqnarray}
  &&{ E}_{N,k-\frac12}(t) =\mathcal C_{N,\nu} \sqrt{t}\ \e^{-Nt/2} 
  \text{U}\left(\frac{N+2m+1}{2},\frac{3}{2},\frac{t}{2}\right)\nn\\
  &&\times\Pf
\left[ \sum_{j=0}^{\frac{N}{2}+m-1} \frac{R^{a}_{2j+1}\left(-t,t\right)R^{b}_{2j}\left(-t,t\right)-(a\leftrightarrow b)}{r_j(t)}\right]_{a,b=0}^{k-1}
  \label{eq:GapPfaffEvenAlpha}
\end{eqnarray}
For $k\!=\!2m-1$ odd 
the last row (and column) inside the Pfaffian is replaced by
$(-)R^{b(a)}_{N+k-2}(-t,t)/r_{N/2+m-1}(t)$, 
respectively (for $N$ odd cf. \cite{AKWW}). The known $t$-independent constant  $\mathcal C_{N,\nu}$ is suppressed for simplicity, it ensures ${ E}_{N,k-\frac12}(t=0)=1$. 
\begin{figure}[t]
  \centering
   \includegraphics[width=84mm]{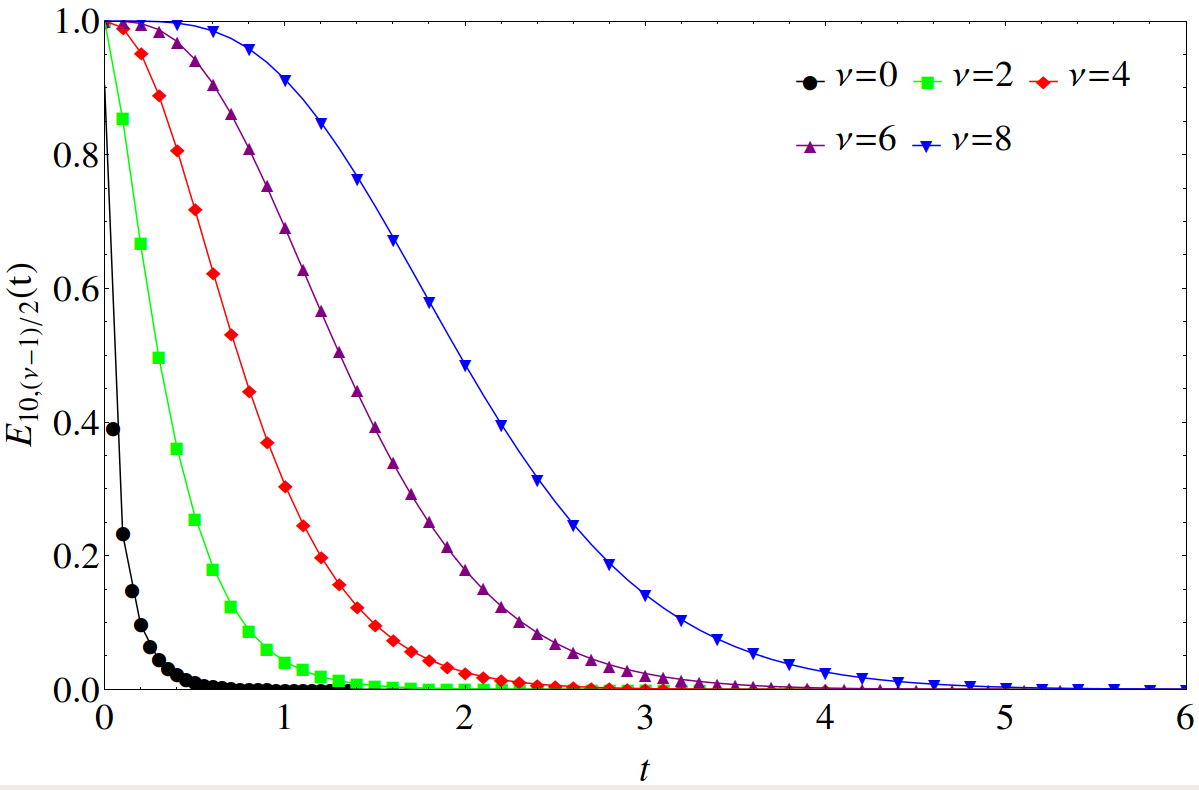}
  \caption{The gap probability 
${ E}_{N,(\nu-1)/2}(t)$ (straight lines) for finite $N=10$ and $\nu=0,2,4,6,8$ (from left to right) vs. numerical simulations (symbols) of $40000$ realisations of Wishart matrices, with $C=\eins_N$.
}
  \label{EN10nu0246}
\end{figure}

Eq.~(\ref{eq:GapPfaffEvenAlpha}) is our first main result. A similar answer can be obtained for ${ P}_{N,\ga}(t)$ for even $\nu$, given in terms of skew-orthogonal polynomials with respect to the weight $x\,w(x;t)$. This provides an explicit integrable Pfaffian structure for both  ${ E}_{N,\ga}(t)$ and ${ P}_{N,\ga}(t)$.
It  extends the odd $\nu$ result for  ${ P}_{N,\ga}(t)$ in \cite{NF98} 
which is given by a Pfaffian determinant as well, but with a different kernel.

For illustration we  give two examples. For $\nu=0$ the Pfaffian in Eq.~(\ref{eq:GapPfaffEvenAlpha}) is absent, 
\be 
\hspace{-0.01cm} { E}_{N,-\frac12}(t) =
\frac{(N-1)!\sqrt{t}\ \e^{-Nt/2}  }{2^{N-1/2}\Gamma(N/2)}
\text{U}\left(\frac{N+1}{2},\frac{3}{2},\frac{t}{2}\right),
\label{Enu0}
\ee
whereas for $\nu=2$ 
the kernel is absent, and only the polynomial (\ref{eq:DifferentiatedEvenPolynomials}) with $a=0$ contributes,
\bea 
{ E}_{N,+\frac12}(t) &=&
\frac{\Gamma\left(\frac{N+1}{2}\right)\sqrt{t}\ \e^{-Nt/2}  }{(-1)^N\sqrt{2\pi}N!}
\label{Enu2}\\
&&\times\left[\text{U}_N(t){L}_N^{(1)}(-t)-\text{U}^\prime_N(t){L}_N^{(0)}(-t)\right].
\nn
\eea
Eqs.~(\ref {Enu0}) and (\ref {Enu2}) are compared to numerical simulations
in Fig.~\ref{EN10nu0246}. They can be matched with the finite $N$ results of \cite{Edelman}  for $\nu= 0, 2$, after differentiating them and using identities for the Tricomi function \cite{NIST}.

\begin{figure}[t]
  \centering
  \includegraphics[width=0.48\textwidth]{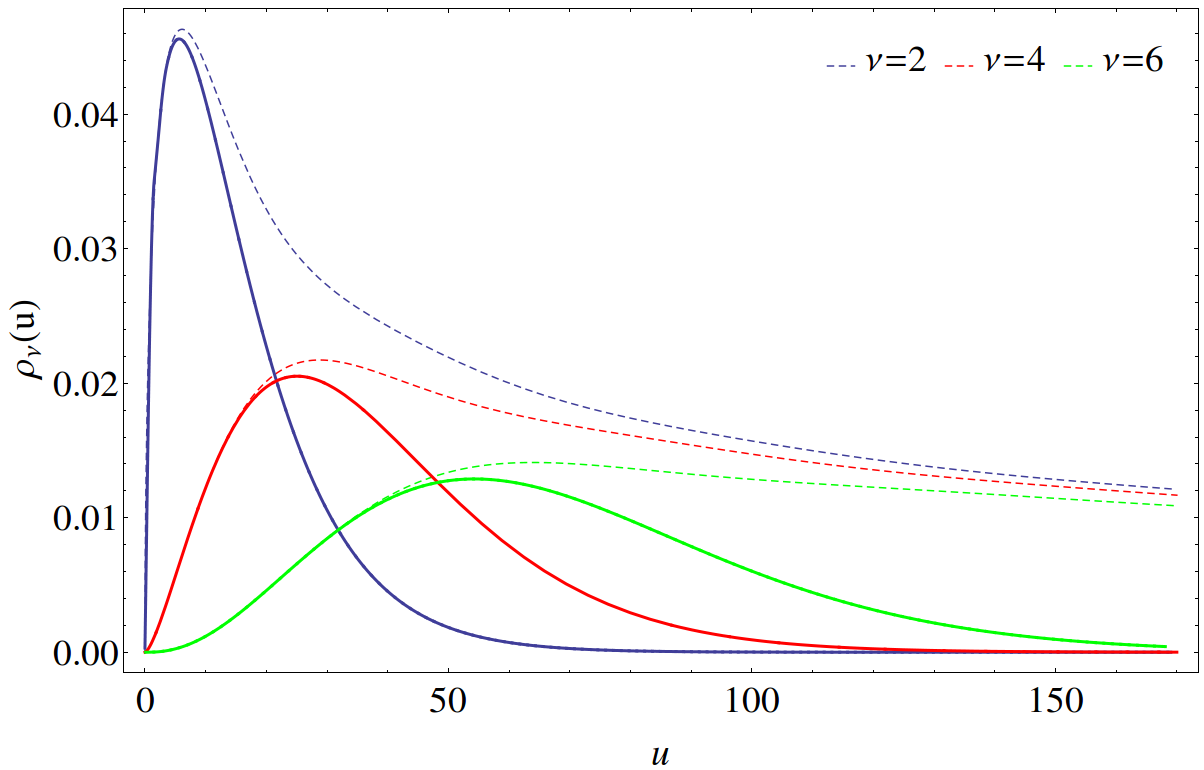}
  \caption{The microscopic density $\rho (u)$ (30) (dashed lines) vs. the corresponding 
smallest eigenvalue distribution ${\cal P}_{(\nu-1)/2}(u)$ (straight lines) for $\nu=2,4,6$ (from left to right). 
The smallest eigenvalue nicely follows the density for all $\nu$.}
  \label{Prhonu246}
\end{figure}
\textbf{Microscopic large $N$ limit.} 
We turn to the large $N$ limit keeping $\nu$ fixed, referred to as hard edge limit. It is particularly important as the limiting density correlation functions are universal for non-Gaussian weight functions for any integer $\nu$ \cite{Klein}. Because the gap probability can be expressed in terms of the limiting universal kernel \cite{Peter06} (see Eq.~(\ref{rho}) for the corresponding density),  
its universality carries over to the distribution of the smallest eigenvalue. Moreover, in \cite{WG} it was shown for {\it both} $\nu$ even and odd, without explicitly calculating the distributions, that the presence of a nontrivial correlation matrix in Eq.~(\ref{measure})
does not change the limiting smallest eigenvalue distribution when the spectrum of $C$ has a finite distance to the origin.

The limiting gap probability and smallest eigenvalue distribution are defined as
\be
{\cal E}_\ga(u)\equiv \lim_{N\to\infty} E_{N,\ga}\left(t=\frac{u}{4N}\right) ,\ \frac{\partial}{\partial u}{\cal E}_\ga(u)=-{\cal P}_\ga(u)~.
\label{limE}
\ee
In view of Eq.~(\ref{eq:GapPfaffEvenAlpha}) we need the following asymptotic limit of the hypergeometric function,
\be
\label{eq:MicroscopicLimitU}
\text{U}\left(aN+c,b,\frac{u}{8N}\right)\approx \frac{2\left(N^2{8a}/{u}\right)^{(b-1)/2}}{\Gamma\left(aN +c\right)}\text{K}_{b-1}\left(\sqrt{\frac{au}{4}}\right).
\ee
For half integer index the modified Bessel function of second kind simplifies, \textit{e.g.} for $b=1/2,3/2,5/2$
\be
K_{\pm\frac12}(z)=\sqrt{\frac{\pi}{2z}}\ \e^{-z},\ K_{\frac32}(z)=(1-z^{-1})K_{\frac12}(z)\ .
\ee
Inside the Pfaffian (\ref{eq:GapPfaffEvenAlpha}) the sum is replaced by an integral, $\sum_{j}\to\frac{N}{2}\int_0^1dx$, with $j=Nx/2$.
The limiting skew-orthogonal polynomials follow from Eq.~(\ref{eq:MicroscopicLimitU}) together with the standard Laguerre asymptotic in terms of modified Bessel functions of the first kind, see e.g. \cite{NIST}. This leads to the following limiting kernel inside the Pfaffian (\ref{eq:GapPfaffEvenAlpha}), independently of $N$ being even or odd,
\bea
&&\kappa_{ab}(u)\equiv\int_0^u\frac{dz}{u}z^{(a+b)/2}
\left[2(b-a)\text{I}_{a}(\sqrt{z})\text{I}_{b}(\sqrt{z})
\right.
\label{limkernel}\\
&&\left.+(2b+1) 
\text{I}_{a+1}(\sqrt{z})\text{I}_{b}(\sqrt{z})-
(2a+1) \text{I}_{b+1}(\sqrt{z})\text{I}_{a}(\sqrt{z})
\right].\nn
\eea
The final answer for the limiting gap probability reads
\be
{\cal E}_{k-1/2}(u)=C_e\ \e^{-\sqrt{u}/2-u/8}
\Pf\left[\kappa_{ab}(u)\right]_{a,b=0}^{k-1}
\label{limEeven}
\ee
for $\nu=2k$ with $k=2m$ even and 
\bea
&&{\cal E}_{k-1/2}(u)=C_o\ \e^{-\sqrt{u}/2-u/8}
\label{limEodd}\\
&&\times\Pf
\left[
\begin{array}{r}
\kappa_{ab}(u)\ \ \ -u^{a/2}\left[ \text{I}_{a+1}\left(\sqrt{u}\right)+\text{I}_{a}\left(\sqrt{u}\right)\right]\\
u^{b/2}\left[ \text{I}_{b+1}\left(\sqrt{u}\right)+\text{I}_{b}\left(\sqrt{u}\right)\right]
\ \ \ \ \ \ \ \ \ \ \ \ \ \ 0\\
\end{array}
\right]_{a,b=0}^{k-1}
\nn
\eea
for $k=2m-1$ odd. We suppress the known $u$-independent normalisation constants  $C_{e/o}$.
\begin{figure}[t]
  \centering
  \includegraphics[width=0.48\textwidth]{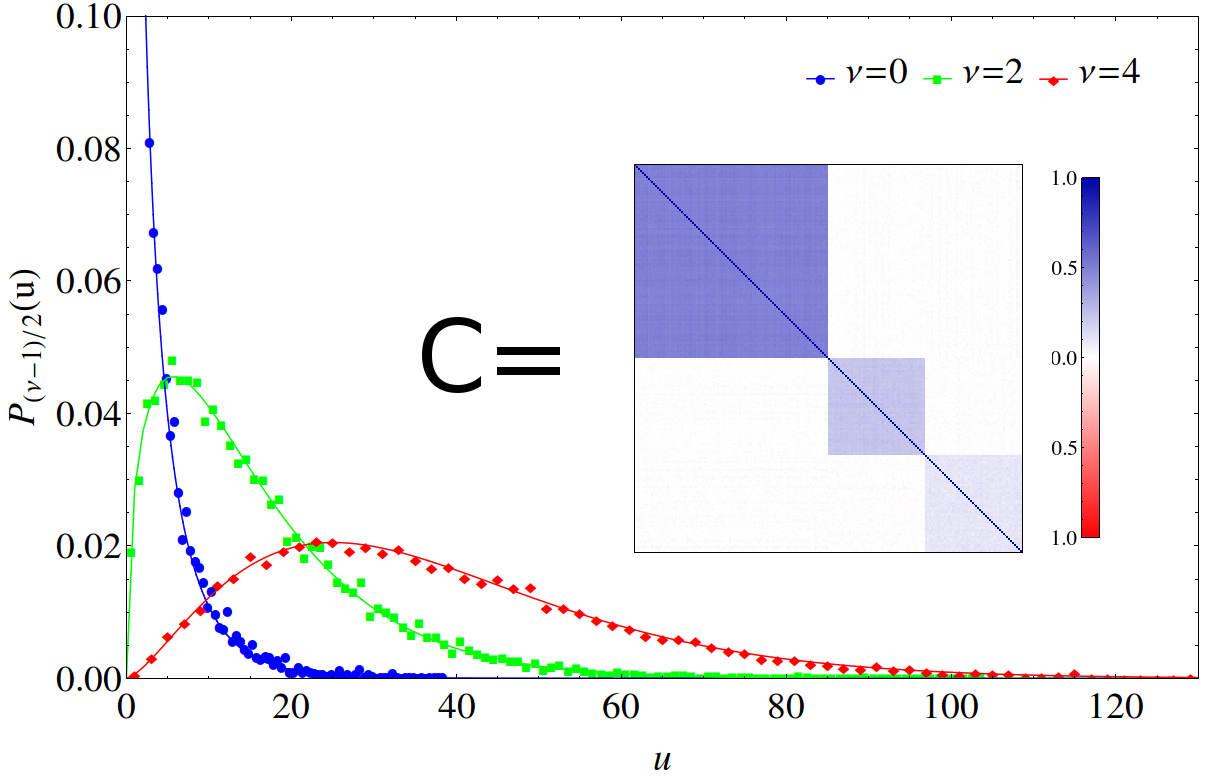}
  \caption{The microscopic 
smallest eigenvalue distribution ${\cal P}_{(\nu-1)/2}(u)$ (straight lines) for $\nu=0,2,4$ (from left to right) vs. numerical simulations (symbols) of $10000$ realisations of matrices with $N=200$ and correlation matrix $C\neq\eins_N$ as indicated in the inset.
}
  \label{PCnu024}
\end{figure}
The corresponding limiting result for the smallest eigenvalue distribution is
\be
{\cal P}_{k-1/2}(u)=\widehat{C}_eu^k(1+2/\sqrt{u})\, \e^{-\sqrt{u}/2-u/8}
\Pf\left[\widehat\kappa_{ab}(u)\right]_{a,b=0}^{k-1}
\label{limPeven}
\ee
for $\nu=2k$ with $k=2m$ even, and 
\bea
&&{\cal P}_{k-1/2}(u)=\widehat{C}_ou^k(1+2/\sqrt{u})\, \e^{-\sqrt{u}/2-u/8}
\label{limPodd}\\
&&\times\Pf
\left[
\begin{array}{r}
\widehat\kappa_{ab}(u)\ \ \ -\frac{ \text{I}_{a+2}\left(\sqrt{u}\right)+\frac{\sqrt{u}}{2+\sqrt{u}}
\text{I}_{a+3}\left(\sqrt{u}\right)}{u^{(a+2)/2}}
\\
\frac{ \text{I}_{b+2}\left(\sqrt{u}\right)+\frac{\sqrt{u}}{2+\sqrt{u}}\text{I}_{b+3}\left(\sqrt{u}\right)}{u^{(b+2)/2}}
\ \ \ \ \ \ \ \ \ \ \ \ \ \ 0\\
\end{array}
\right]_{a,b=0}^{k-1}\nn
\eea
for $k=2m-1$ odd, suppressing again the $u$-independent normalisation constants  $\widehat{C}_{e/o}$. Here $\widehat\kappa_{ab}(u)$ is the limiting kernel for the skew-orthogonal polynomials with respect to $x\,w(x;t)$ which is of a similar structure as Eq.~(\ref{limkernel}). For $\nu=0,2$ the results (\ref{limPeven}) and (\ref{limPodd}) were known from \cite{Peter93}, \cite{AV}, respectively.

Eqs.~(\ref{limEeven}) - (\ref{limPodd}) constitute our second main result and are universal. 
In Fig. \ref{Prhonu246}
they are compared to the universal microscopic density \cite{Jac,FNH}
valid for all $\nu$-values
\bea
\rho_\nu(u)&=&\frac{1}{4}\left( 
J_\nu(\sqrt{u})^2-J_{\nu-1}(\sqrt{u})J_{\nu+1}(\sqrt{u})\right)
\nn\\
&&+\frac{1}{4\sqrt{u}} J_\nu(\sqrt{u})\left(1-\int_0^{\sqrt{u}}dsJ_\nu(s) 
\right)~.
\label{rho}
\eea
We further illustrate the universality of our results by comparing to numerical simulations with a nontrivial correlation matrix $C$ for large $N$, see Fig. \ref{PCnu024}.

\textbf{Conclusions and outlook.}  
We have computed closed expressions for the distribution of the smallest non-zero eigenvalue and its integral, the gap probability, for rectangular $N\times(N+\nu)$ real Wishart matrices 
with $\nu$ even, both for finite $N$ and in the universal microscopic large $N$ limit. 
They only depend on a single kernel instead of three different ones for the density correlation functions and are thus much simpler than these known results.
We confirm our findings by numerical simulations even including a nontrivial correlation matrix $C$.
This completes the calculation of all eigenvalue correlation functions in this classical ensemble of random matrices and shows its integrable structure.
Furthermore, our finite $N$ results allow to analyse deviations from the 
universal large $N$ limit, as was very recently proposed in \cite{EGP} for the
chGUE. 
\\[-3mm]

\textbf{Acknowledgements.} We thank the Son\-der\-for\-schungs\-be\-reich TR12 (G.A., T.G. and T.W.) and the Alexander von Humboldt-Foundation (M.K.) for support.


\begin{thebibliography}{XX}
\bibitem{footnote1}
For $N$ odd an additional condition has to be imposed, $\int_0^\infty dx R_j(x,t)=\delta_{j,N-1}$, cf. \cite{Mehta} .
\end{thebibliography}
\end{document}